\documentclass[11pt]{article}
\usepackage{graphicx}
\usepackage[table]{xcolor}
\usepackage{amsmath}
\usepackage{array,booktabs,siunitx}
\usepackage[utf8]{inputenc}
\usepackage{multirow}
\usepackage{fmtcount} 
\usepackage{array,multirow}
\newcolumntype{C}[1]{>{\centering\arraybackslash}m{#1}}
\usepackage{caption}
\usepackage{hyperref}
\usepackage{fancyhdr}

\usepackage{titling}

\usepackage{amsmath,amssymb}
\usepackage{slashed}
\usepackage{xcolor} 
\usepackage{graphicx}
\usepackage{dcolumn} 
\usepackage{bm} 
\usepackage{epsfig}
\usepackage{epstopdf}
\usepackage{grffile}
\usepackage{color}
\usepackage{colordvi}
\usepackage{amsmath,amssymb}
\usepackage{rotating}
\usepackage{lscape}
\usepackage{cite}
\usepackage{float}
\usepackage{hyperref}
\usepackage{url}
\usepackage{graphicx}
\usepackage{color}
\usepackage{amssymb}
\usepackage{amsmath}
\usepackage{mathtools}
\usepackage{bm}	
\usepackage{flexisym}
\usepackage{amssymb}
\usepackage{physics}
\usepackage{mathrsfs}
\usepackage{simplewick}
\usepackage{tikz}
\usetikzlibrary{arrows,shapes}
\usetikzlibrary{trees}
\usetikzlibrary{matrix,arrows} 			
\usetikzlibrary{positioning}			
\usetikzlibrary{calc,through}			
\usetikzlibrary{decorations.pathreplacing}  
\usepackage{pgffor}						
\usetikzlibrary{decorations.pathmorphing}
\usetikzlibrary{decorations.markings}
\tikzset{
	vector/.style={decorate, decoration={snake}, draw},
	provector/.style={decorate, decoration={snake,amplitude=2.5pt}, draw},
	antivector/.style={decorate, decoration={snake,amplitude=-2.5pt}, draw},
	fermion/.style={draw=black, postaction={decorate},
		decoration={markings,mark=at position .55 with {\arrow[draw=black]{>}}}},
	fermionbar/.style={draw=black, postaction={decorate},
		decoration={markings,mark=at position .55 with {\arrow[draw=black]{<}}}},
	fermionnoarrow/.style={draw=black},
	gluon/.style={decorate, draw=black,
		decoration={coil,amplitude=4pt, segment length=5pt}},
	scalar/.style={dashed,draw=black, postaction={decorate},
		decoration={markings,mark=at position .55 with {\arrow[draw=black]{>}}}},
	scalarbar/.style={dashed,draw=black, postaction={decorate},
		decoration={markings,mark=at position .55 with {\arrow[draw=black]{<}}}},
	scalarnoarrow/.style={dashed,draw=black},
	electron/.style={draw=black, postaction={decorate},
		decoration={markings,mark=at position .55 with {\arrow[draw=black]{>}}}},
	bigvector/.style={decorate, decoration={snake,amplitude=4pt}, draw},
}
\tikzstyle{block} = [draw, rectangle, minimum height=3em, minimum width=6em]
\usepackage{hyperref}

\usepackage{titling}
\usepackage{subcaption}
\newcommand{\subtitle}[1]{%
	\posttitle{%
		\par\end{center}
	\begin{center}\large#1\end{center}
	\vskip0.5em}%
}

\def\Re{{\cal R \mskip-4mu \lower.1ex \hbox{\it e}\,}}
\def\Im{{\cal I \mskip-5mu \lower.1ex \hbox{\it m}\,}}

\def\tev{\,{\ifmmode\mathrm {TeV}\else TeV\fi}}
\def\gev{\,{\ifmmode\mathrm {GeV}\else GeV\fi}}
\def\mev{\,{\ifmmode\mathrm {MeV}\else MeV\fi}}
\def\to{\rightarrow}

\begin{document}

\begin{center}

\vspace*{15mm}
\vspace{1cm}
{\Large \bf Machine Learning Approaches to Top Quark Flavor-Changing Four-Fermion Interactions in Trilepton Signals at the LHC}

\vspace{1cm}

{\bf Meisam Ghasemi Bostanabad and Mojtaba Mohammadi Najafabadi }\\
{\small\sl 
School of Particles and Accelerators, Institute for Research in Fundamental Sciences (IPM) P.O. Box 19395-5531, Tehran, Iran\\
 }\vspace*{.2cm}
\end{center}

\vspace*{.2cm}

\vspace*{10mm}

\begin{abstract}\label{abstract}
We explore the top quark flavor-changing 4-Fermi interactions ($tuee$ and $tcee$) with scalar, vector, 
and tensor structures using machine learning models to analyze tri-lepton processes at the LHC.
The study is performed using $t\bar{t}$ and $tW$ processes, where a top quark decays into $u/c+e^{+}+e^{-}$.
 The analysis incorporates both reducible and irreducible 
backgrounds while accounting for realistic detector effects. The dominant backgrounds for these 
trilepton signatures arise from $t\bar{t}$ production, single top quark production in association with $V$, and $VV$ 
production (where $V = W, Z$). These backgrounds are significantly reduced using machine learning-based classification models, 
which optimize event selection and improve signal sensitivity.
 For an integrated luminosity of 3000 fb$^{-1}$ at the LHC, we find that the expected $95\%$ confidence level (CL) limits on the 
 scale of 4-Fermi FCNC interactions reach $\Lambda \leq 5.5$ TeV for $tuee$ and $\Lambda \leq 5.7$ TeV for $tcee$ in the $t\bar{t}$ channel, 
 and $\Lambda \leq 1.9$ TeV ($tuee$) and $\Lambda \leq 2.0$ TeV ($tcee$) in the $tW$ channel.
 We also provide an interpretation of our EFT analysis in the context of a specific $Z'$ model, 
 illustrating how the derived constraints translate into bounds on the parameter space of a heavy 
 neutral gauge boson mediating flavor-changing interactions.
  \end{abstract}
  \vspace{4cm}


\newpage

\section{Introduction}
\label{sec:introduction}

With its large mass and strong Higgs boson coupling, the top quark is key to probing flavor physics
 beyond the Standard Model (SM), particularly in searches for new interactions violating flavor conservation.
Its substantial mass makes it particularly sensitive to 
various forms of new physics, especially in relation to flavor-changing interactions and 
CP violation. In many scenarios of new physics, flavor-changing neutral current (FCNC) 
effects involving the top quark are enhanced due to the top quark’s mass relative to the 
scale of the new physics. As a result, studying top quark interactions provides a powerful 
probe into potential new flavor dynamics beyond the SM.
In the SM, FCNCs
 are forbidden at tree level due to the Glashow-Iliopoulos-Maiani (GIM) mechanism \cite{Glashow:1970gm}, which ensures their suppression. 
 Consequently, FCNC processes can only occur at higher orders through loop-level diagrams and are highly suppressed, 
 leading to branching fractions far below current experimental sensitivities.
 In particular, transitions like $t \to u$ and $t \to c$, either in top decays or production processes, 
 are extraordinarily rare under SM predictions, with branching ratios far below current experimental sensitivities.
 
 In the SM, the branching ratios for FCNC top quark decays \( t\to u/c +\ell^{+}+\ell^{-} \), 
 are exceedingly small and is at the order of $10^{-14,-13}$ \cite{m1, m11}. 
In the SM, the loop amplitudes for these decays are dominated by contributions from down-type quarks, 
with the bottom quark (\( b \)) playing the most significant role. Consequently, the scale of these 
amplitudes is set by the bottom quark mass \( m_b \). For instance, the partial widths for \( t\to cZ \) 
decays can be expressed as \cite{m1, m11}:
\[
\Gamma(t\to Z+c) \sim \left|V_{tb}^{*} V_{cb}\right|^2 \alpha_{\text{em}} \times g^4 \times m_t (m_b/m_W)^4 \times \mathcal{F},
\]
where \( \alpha_{\text{em}} \)  represents fine structure constant, \( g \) is the weak coupling constant, and $V_{tb}, V_{cb}$
are CKM matrix elements. $\mathcal{F}  \sim \left(1 - m_Z^2/m_t^2 \right)^2$ 
is a phase space and polarization factor obtained by neglecting the charm quark mass \( m_c \). 
The factor \( \left(m_b/m_W\right)^4 \) arises from the GIM mechanism, 
which introduces an additional suppression beyond naive expectations based on 
dimensional analysis, power counting, and CKM matrix elements. 
This fourth power of the mass ratio is responsible for the extreme 
suppression of these decays, as the GIM mechanism cancels out leading-order 
contributions and leaves only terms proportional to the mass differences of the 
down-type quarks in the loop. 
This suppression of FCNC effects within the SM implies that even the smallest deviation from these expectations, 
either in direct measurements or indirect observables, may provide compelling evidence for new physics beyond 
the SM. Top quark FCNC processes are thus of great interest, as they offer a clean signal for exploring potential NP 
effects that may not be observable in lighter quark sectors.
Theoretical and experimental efforts have been dedicated to the study of top quark FCNCs, both within 
model-independent effective field theory (EFT) frameworks and in the context of specific extensions to the SM, 
such as models involving new heavy gauge bosons, extra dimensions, 
or supersymmetry \cite{m1,m11,m2,m3,m4,m5 ,e1, e2, e3, e31, e4 ,e5 ,e6, e7, e8, e9, e10, e101,e11, e12, e13, e14, e15, e16, e17}. 

In this study, we investigate the FCNC processes described by 
the effective Lagrangian which will be described in the next section. 
Our analysis probes two key channels: top pair production, 
where one of the top quarks decays into a light up-type quark ($u$ or $c$) and two same-flavor, 
opposite-sign charged electrons; and single top quark production in association with a $W$ boson, 
where the top quark undergoes a flavor-changing decay as described by the effective Lagrangian. 
In top quark pair production case, the semi-leptonic decay of the other top quark is considered and 
for the $tW$ production process, the leptonic decay of the $W$ is probed.
Therefore, the final state is characterized by a trilepton signature accompanied by missing transverse momentum and additional jets. 
This focus on trilepton final states is strategic, as these states provide a promising environment for observing FCNC processes 
with minimal SM background interference and reduced sensitivity to systematic uncertainties. 
The analysis is further refined by employing advanced machine learning techniques, enhancing signal-to-background discrimination 
through optimized classification algorithms. 
These methods significantly boost sensitivity, enabling a clearer distinction of rare FCNC-driven top quark processes 
from dominant SM backgrounds, thus advancing the precision measurement of these elusive interactions.

The structure of the paper is as follows: 
Section \ref{sec:th} is dedicated to describing the theoretical framework and the flavor-changing 
four-fermion interactions of the top quark, along with their experimental constraints.
Section \ref{sec:generation} provides a detailed
 description of the signal and background process definitions and their generation. 
 In Section \ref{sec:machinelearning}, we present the machine learning algorithms 
 employed to achieve optimal discrimination between signal and background processes. 
 The statistical methods used to derive exclusion limits, along with the final results, 
 are discussed in Section \ref{res}. 
In Section \ref{sec:int}, an interpretation of the EFT analysis in the context of a specific flavor-changing \( Z' \) model
is presented. Lastly, a summary of findings and conclusions is given in Section \ref{summ}.

\section{Top Quark Flavor-Changing Four-Fermion Interactions and Experimental Constraints}
\label{sec:th}

To study flavor-changing four-fermion interactions involving the top quark, a model-independent approach is adopted. 
Such approaches typically focus on parameterizing new physics effects in terms of higher-dimensional operators, 
such as four-Fermi interactions, that modify the SM Lagrangian. 
These operators are suppressed by the new physics scale $\Lambda$, but the large mass of the top quark significantly enhances their potential observability.  
In this study, we adopt the parametrization from Refs.~\cite{p1,p2,p3,p4} for the effective Lagrangian 
describing flavor-changing $t\bar{u}\ell^{+}\ell^{-}$ contact interactions, given by:  
\begin{equation}\label{effLag}
	\mathcal{L}_{t u \ell \ell}=\frac{1}{\Lambda^2_{\ell}} \sum_{i, j=L, R}\left[V_{i j}^{\ell}\left(\bar{\ell} \gamma_\mu P_i \ell\right)\left(\bar{t} \gamma^\mu P_j u\right)+S_{i j}^{\ell}\left(\bar{\ell} P_i \ell\right)\left(\bar{t} P_j u\right)+T_{i j}^{\ell}\left(\bar{\ell} \sigma_{\mu \nu} P_i \ell\right)\left(\bar{t} \sigma_{\mu \nu} P_j u\right)\right],
\end{equation}  
where $u$ represents up- and charm-quarks, $\ell$ represents the electron, muon, and $\tau$
and the different couplings are given by $V_{i j}^{\ell}$ (vector-like), $S_{i j}^{\ell}$ (scalar-like), and $T_{i j}^{\ell}$ (tensor-like).  
In this formulation, the Wilson coefficients $V_{i j}^{\ell}$, $S_{i j}^{\ell}$, $T_{i j}^{\ell}$, and the new physics scale $\Lambda_{\ell}$ 
are assumed to be different for different lepton flavors. This assumption is well-motivated by both theoretical and experimental considerations. 
Many BSM scenarios predict non-universal couplings across different lepton generations, potentially leading to 
lepton flavor universality violation, as suggested by anomalies observed in B physics. 
Furthermore, experimental constraints on FCNCs involving different 
leptons vary significantly. While electron-related FCNCs are tightly constrained by 
high-precision low-energy experiments, those involving muons and $\tau$s are subject to 
comparatively looser bounds, allowing for different new physics scales across lepton flavors \cite{t1,t2,t3}.
The decay rate for the three-body process \( t \to u + \ell^{+}  + \ell^{-}  \) is given by:
\begin{equation}
    \Gamma(t \to u/c + \ell^{+} + \ell^{-} ) = \frac{m_{t}^{5}}{6144 \pi^3  \Lambda_{\ell}^4} \sum_{i, j=L, R} \left[ |V_{ij}^\ell|^2 \, \mathcal{I}_V + |S_{ij}^\ell|^2 \, \mathcal{I}_S + |T_{ij}^\ell|^2 \, \mathcal{I}_T \right],
\end{equation}
where  \( \mathcal{I}_V  = 4\), \( \mathcal{I}_S  = 1\), and \( \mathcal{I}_T =48 \). The interference terms vanish in this approximation, 
as they are proportional to \( m_\ell \) or \( m_u \).
The tensor coupling \( T_{ij}^\ell \) dominates  with respect to the vector and scalar type couplings
which is mainly due to its momentum-dependent terms and Lorentz structure, 
which enhance its contribution when integrating over the phase space.

In previous studies, the flavor-changing 4-Fermi interactions involving top quarks, 
such as $tu\ell\ell$ and $tc\ell\ell$ described by Lagrangian presented in Eq.\ref{effLag}, 
have been extensively explored as potential signatures of NP beyond the SM \cite{p3,p4, p5,p6, p7, p8, p9,p10}. 
These interactions are hypothesized to arise from high-energy processes involving heavy mediators, 
such as new scalar, vector, or tensor fields, which induce higher-dimensional operators in effective field theories. 
These FCNC interactions introduced in Eq.\ref{effLag} open up novel production channels at the LHC, 
such as single-top production accompanied by a dilepton pair ($pp \rightarrow t\ell\ell$) or with an 
additional light jet ($pp \rightarrow t\ell\ell + j$). These processes can manifest in distinct experimental signatures, 
notably in the form of dilepton + $b$-jet events ($pp \rightarrow \ell\ell + b$-jet + X) and tri-lepton final 
states ($pp \rightarrow \ell\ell\ell' + X$), where two of the leptons are opposite-sign same-flavor (OSSF) pairs \cite{p4}.

LEP2 analyses set limits on 
the $tu\ell\ell$ and $tc\ell\ell$ four-Fermi interactions, 
constraining the new physics scale $\Lambda_{e}$ to approximately 1 TeV \cite{p5,p6}. 
However, the LHC has the potential to significantly improve these limits, providing a more sensitive probe of such interactions.
For example, analyses of the dimuon channel ($pp \rightarrow \mu\mu + b$-jet + X) 
with existing LHC data corresponding to 140 fb$^{-1}$ have provided that $95\%$ CL 
bounds on the NP scale reach $\Lambda_{\mu} \leq 5$ TeV for tensor-like interactions and 
$\Lambda_{\mu} \leq 3.2$ TeV for vector-like interactions for the $tu\mu\mu$ interaction \cite{p4}.

The precise measurement performed by the MOLLER experiment \cite{moller} places 
stringent constraints on new physics, particularly on four-fermion contact interactions 
that contribute to electron-electron scattering. 
From this measurement, the following limits have been 
derived \cite{e15}: \( \Lambda_{e}/\sqrt{c_{tuee}} > 3.0 \) TeV and \( \Lambda_{e}/\sqrt{c_{tcee}} > 1.7 \) TeV for the \( tuee \) and \( tcee \) 
interactions, respectively. 
Consequently, these constraints impose strong upper limits on the branching ratios of rare top decays, 
specifically \( Br(t \to c+ e^{+} + e^{-}) \lesssim 3 \times 10^{-7} \) and \( Br(t \to u+ e^{+} + e^{-}) \lesssim 3 \times 10^{-8} \).

\begin{figure}[ht] 
  \centering
    \includegraphics[width=0.6\linewidth]{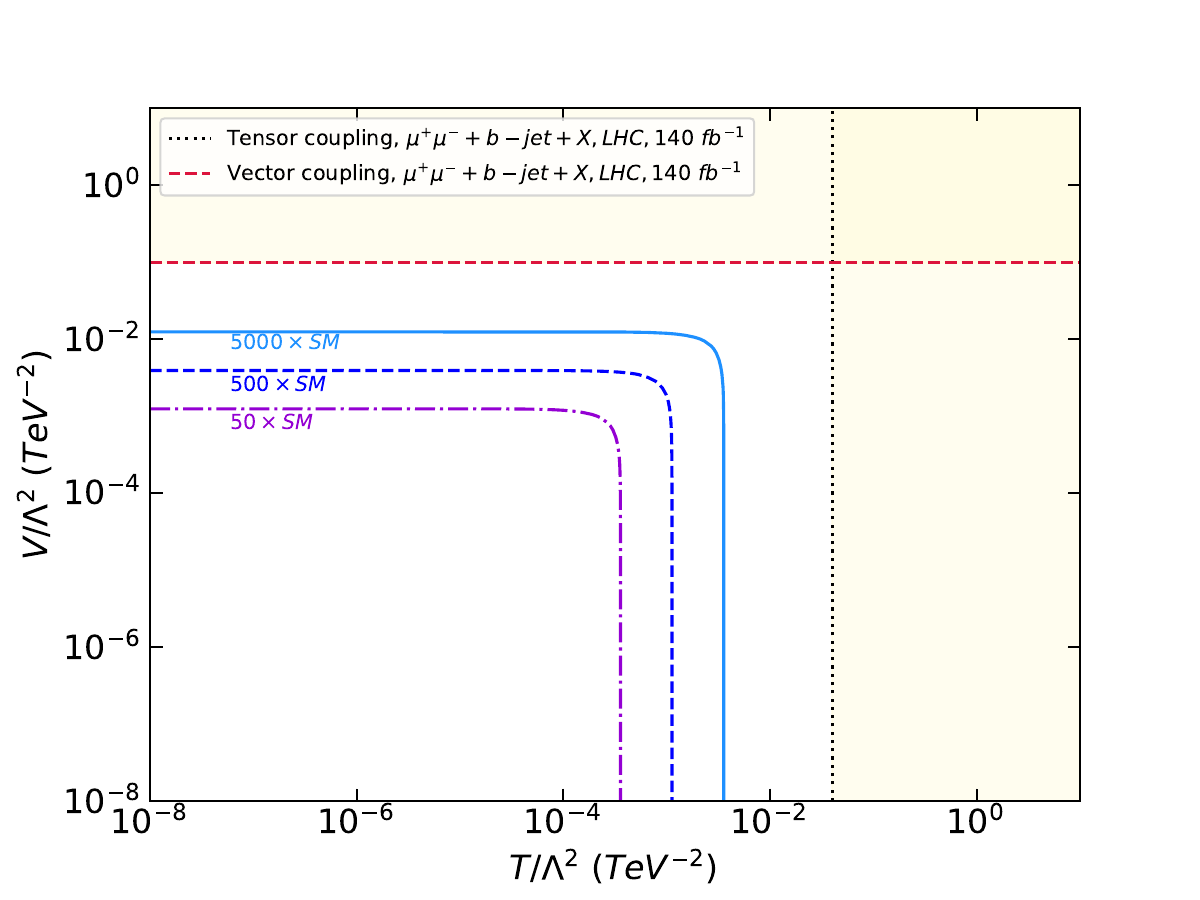}
  \caption{The branching fraction contours for $t \to u + \ell^{+} +  \ell^{-}$, expressed in units of the SM branching fraction, 
  are shown in the plane of vector and tensor coupling strengths. The contour levels are set to 50, 500, and 5000 times the SM value, which is taken to be $2.8\times 10^{-13}$, for illustration.
  The excluded regions for tensor and vector couplings, derived using 140 fb$^{-1}$ of LHC data, are indicated by dotted and dashed lines with a yellow-shaded area \cite{p4}. }
   \label{bbrr}
\end{figure}

Figure~\ref{bbrr} presents the branching fraction contours for $t \to u/c + \ell^{+} + \ell^{-}$, 
expressed in units of the SM branching fraction, in the plane of vector and tensor coupling strengths. 
The contour levels are set to 50, 500, and 5000 times the SM value for illustration, where the SM prediction is taken to be approximately $10^{-13}$. 
As observed, the current experimental limits remain significantly above the SM expectation. 
The excluded regions for the vector and tensor couplings, derived using 140 fb$^{-1}$ of LHC data, 
are shown by dotted and dashed lines with a yellow-shaded region \cite{p4}.

Looking ahead to the High-Luminosity LHC (HL-LHC), with a projected dataset of 3000 fb$^{-1}$, 
the discovery potential is even more promising. The anticipated $95\%$ CL bounds on the 
scale of the $tu\mu\mu$ tensor (vector) interactions are expected to reach $\Lambda_{\mu} \leq 7.1$ TeV ($4.7$ TeV), 
while for the $tc\mu\mu$ interactions, the bounds are projected to be $\Lambda_{\mu} \leq 2.4$ TeV ($1.5$ TeV) 
for tensor (vector) couplings \cite{p4}. These enhanced sensitivities demonstrate the LHC’s capability to probe the 
high-energy regime of new flavor-changing physics and place stringent limits on the possible existence of these 4-Fermi operators.

In a recent study by the CMS experiment, searches for charged-lepton flavor violation (CLFV)
 in top quark production and decay were reported, focusing on potential 4-Fermi interactions 
 involving $tue\mu$ and $tce\mu$ \cite{cms1}. These interactions, which are forbidden in the SM, 
 serve as a sensitive probe for new physics beyond the SM, particularly in scenarios involving FCNCs. 
 The analysis utilized data from proton-proton collisions at a center-of-mass energy of 13 TeV, 
 corresponding to an integrated luminosity of 138 fb$^{-1}$, collected by the CMS detector at the LHC.
Events were selected based on the presence of an oppositely charged electron-muon pair and at least one b-tagged jet, 
which is a characteristic signature of top quark decays. The analysis employed an EFT framework to parameterize the 
lepton flavor-violating interactions of the top quark. Both the production and decay modes of the top quark mediated by
 these effective operators were included in the search, enabling a comprehensive investigation of CLFV effects in the top sector.
For the $tue\mu$ coupling, the $95\%$ CL lower limits on $\Lambda/\sqrt{c}$ are found to be 6.7, 4.8, and 3.3 TeV for the tensor, vector, 
and scalar couplings, respectively, where $c$ represents the corresponding Wilson coefficient. Similarly, for the $tce\mu$ coupling, 
the $95\%$ CL lower limits on $\Lambda/\sqrt{c}$ are measured to be 3.4, 2.4, and 1.6 TeV for the tensor, vector, and scalar couplings, respectively.

\section{Simulated Samples and Selection Criteria for FCNC Events}
\label{sec:generation}

\begin{figure}[ht]
  \centering
    \includegraphics[width=0.6\linewidth]{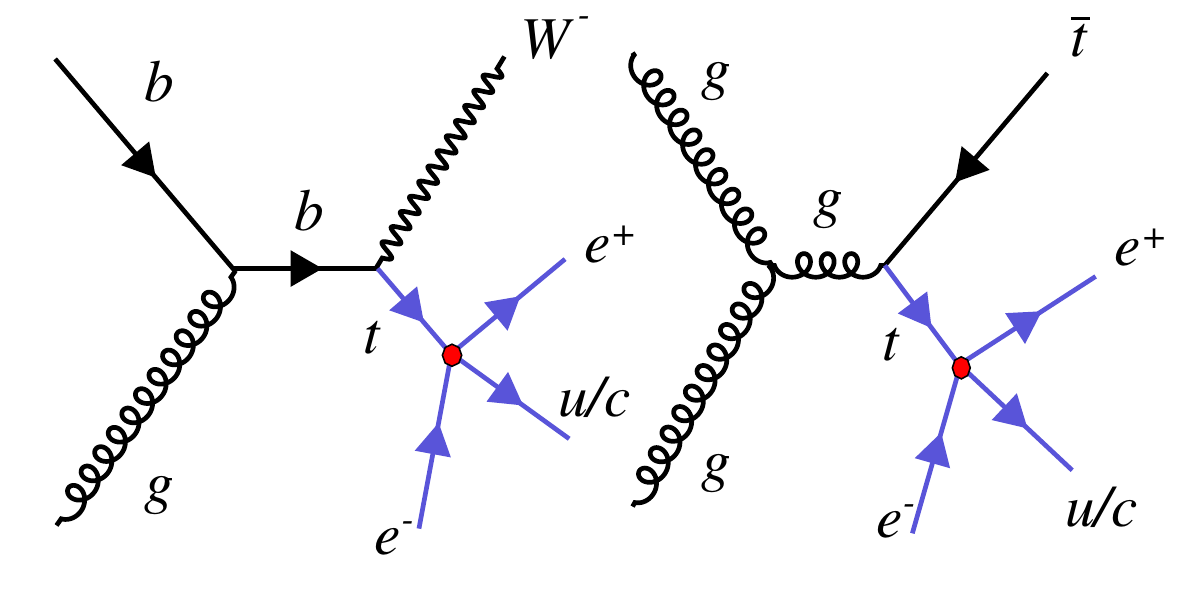}
  \caption{Representative Feynman diagrams for the $pp \rightarrow tW$ (left) and $pp \rightarrow t\bar{t}$ (right) signal events. 
  The red circle signifies the FCNC interaction of the top quark, which is the focus of this search. }
   \label{figfeyn}
\end{figure}

In this analysis, two signal processes, $t\bar{t}$ and $tW$, are examined to probe the flavor-changing couplings 
of the top quark through the decays $t \to u + e^{+} + e^{-}$ and $t \to c + e^{+} + e^{-}$. 
The study of FCNC couplings to up and charm quarks is performed independently for each channel. 
In the $t\bar{t}$ process, one top quark undergoes an FCNC decay into a light up or charm quark accompanied by two opposite-sign (OS) 
electrons, while the other top quark decays conventionally into a $W$ boson and a $b$-quark. For both the $t\bar{t}$ and $tW$ signals, 
the $W$ boson decays leptonically, contributing an additional electron in the final state. 
Consequently, signal events are characterized by the presence of three electrons (two OS electrons from the FCNC decay and one from the $W$ decay), 
along with at least a light-flavor jet (from the up or charm quark) and exactly one $b$-jet in the case of $t\bar{t}$ production. 
 The Feynman diagrams for these signal processes are shown in Figure \ref{figfeyn}, illustrating the expected final states for each channel.
 
For both processes, the expected number of signal events, denoted as $\mathcal{N}_{S(tW)}$ and $\mathcal{N}_{S(t\bar{t})}$, is given by:
\begin{eqnarray}
\mathcal{N}_{S(tW, t\bar{t})} = \sigma_{(tW, t\bar{t})} \times \mathcal{B}(t \rightarrow u/c+e^{+}+e^{-}) \times \epsilon_{(tW, t\bar{t})} \times \mathcal{L},
\end{eqnarray}
where $\sigma_{(tW, t\bar{t})}$ represents the production cross-section of the respective process, 
$\mathcal{B}(t \rightarrow u/c+e^{+}+e^{-})$ denotes the branching fraction of the FCNC decay, $\epsilon_{(tW, t\bar{t})}$ 
accounts for the overall selection efficiency specific to the $tW$ and $t\bar{t}$ channels, 
and $\mathcal{L}$ is the integrated luminosity of the dataset, taken as 3000 fb$^{-1}$. 
The efficiency factor $\epsilon_{(tW, t\bar{t})}$ incorporates multiple contributions, 
including the impact of event selection criteria, detector acceptance, 
and reconstruction performance, which are inherently dependent on the interaction type.
Given that the reconstruction efficiency of leptons varies with their flavor, $\epsilon_{(tW, t\bar{t})}$ is lepton-flavor dependent. 
Electrons generally experience lower detection efficiency compared to muons due to their higher probability of undergoing 
bremsstrahlung and energy loss in the detector material, leading to broader energy deposits and more stringent isolation requirements. 
Additionally, electron identification is more sensitive to detector resolution effects, particularly at high pseudorapidity, 
which further reduces efficiency relative to muons. 
In contrast, muons, which traverse the entire detector and leave distinct signatures in the muon spectrometer, 
are less affected by such effects, resulting in a comparatively higher reconstruction efficiency. 
Consequently, the sensitivity in the electron channel is expected to be weaker compared to the muon 
channel when analyzed under similar conditions.

The primary background processes contributing to both signal channels include 
top-antitop ($t\bar{t}$) production, diboson processes such as $ZZ$ and $W^\pm Z$, 
associated top-quark production with vector bosons ($t\bar{t}W^\pm$, $t\bar{t}Z$, $tZq$), 
and four-top-quark ($4t$) production. The dominant SM background for the $t\bar{t}$ signal
 arises primarily from $t\bar{t}$ production, where electron misidentification can occur due to 
 jet misidentification or photon conversion, leading to an irreducible background contribution. 
 In the case of the $tW$ production channel, the most significant backgrounds originate from 
 diboson production ($VV$, where $V = W, Z$) and single-top-quark production in association 
 with a $Z$ boson ($tZq$), with subsequent leptonic decays of the vector bosons. 
The analysis is conducted at the LHC with a center-of-mass energy of 13 TeV, 
assuming an integrated luminosity of 3000 fb$^{-1}$, 
corresponding to the dataset anticipated for the High-Luminosity LHC phase.

The effective Lagrangian described in Eq. \ref{effLag} was implemented into the FeynRules framework \cite{feynrule}, 
generating a Universal Feynman Output (UFO) model \cite{ufo}. This model was then integrated into \texttt{MadGraph5_aMC@NLO} \cite{MG}, 
enabling precise numerical cross-section calculations and event generation. Both signal and background events 
were simulated with \texttt{MadGraph5_aMC@NLO}, followed by parton showering, hadronization, and underlying event simulations performed with \texttt{PYTHIA v8} \cite{PYTHIA}.

All samples were subsequently processed using \texttt{DELPHES 3} \cite{DELPHES}, employing the high-luminosity 
CMS detector configuration to simulate realistic detector effects and to reconstruct final-state objects such as electrons, 
jets, and missing transverse energy. Electron reconstruction relied on isolation criteria within a 
cone of $\Delta R=\sqrt{(\Delta \eta)^{2} + (\Delta \phi)^{2}} = 0.3$, with minimum transverse momentum ($p_{T}$) set to 20 GeV 
and pseudorapidity restricted to $|\eta| < 3$. Jets were reconstructed using the anti-$k_{t}$ clustering algorithm \cite{antikt} 
with a radius parameter of $R=0.4$, implemented through the \texttt{FASTJET} package \cite{fastjet1, fastjet2}. 
Jet candidates were selected with a minimum transverse momentum requirement of \( p_T > 30 \) GeV and a pseudorapidity constraint of \( |\eta| < 3 \). 
In the analysis of the \( t\bar{t} \) and \( tW \) processes, the number of jets was required to be at least two and one, respectively. 
For the \( t\bar{t} \) signal process, exactly one \( b \)-tagged jet was required.

\begin{figure}[h]
  \centering
    \includegraphics[width=0.5\linewidth]{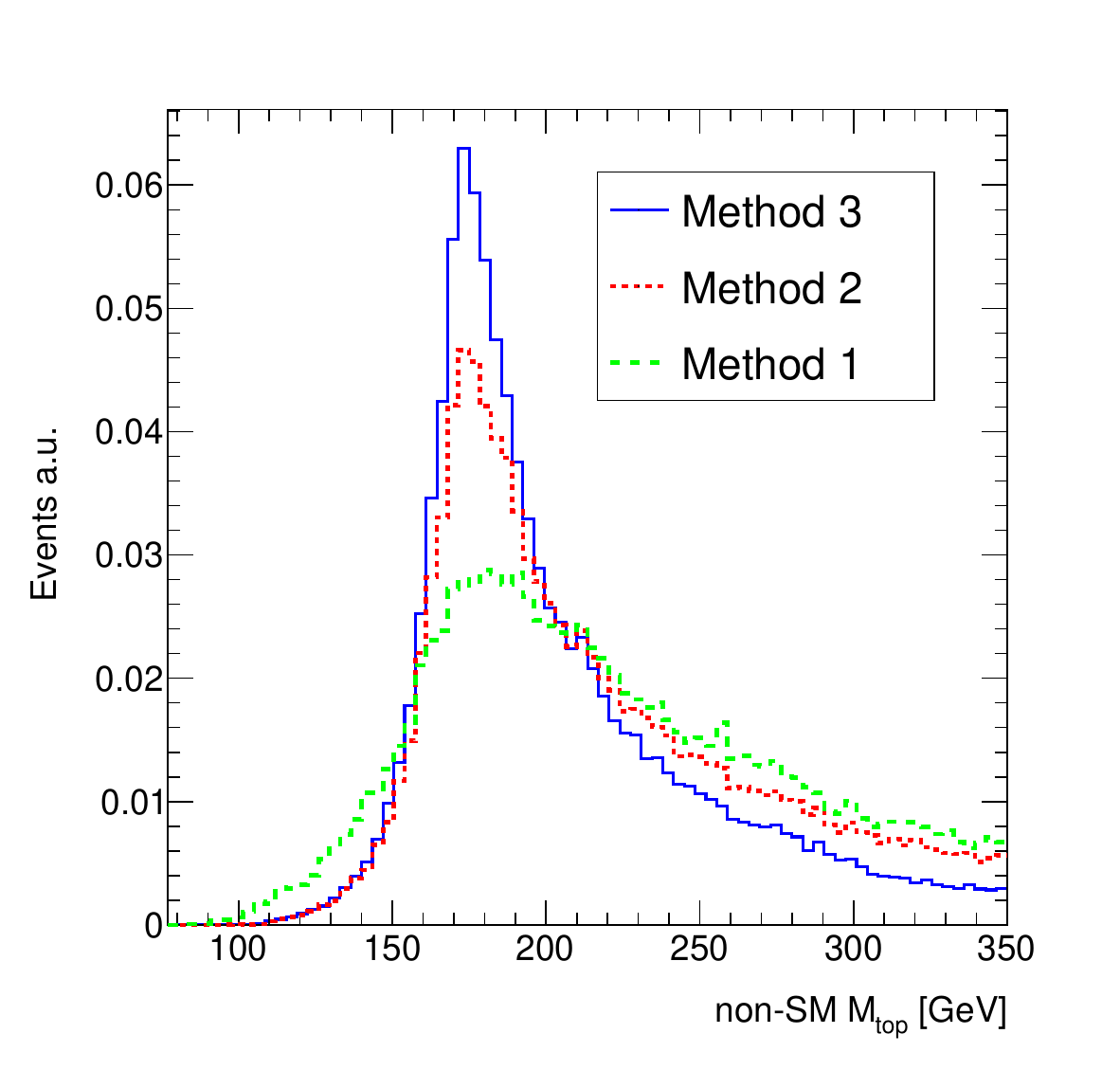}
  \caption{ Top quark mass distribution reconstructed from the FCNC vertex using three algorithms. 
  The green, red, and blue histograms correspond to methods utilizing different combinations of opposite-sign electrons 
  and non-$b$-tagged jets to optimize the reconstruction accuracy. 
  The blue histogram, representing the most comprehensive algorithm, provides the closest approximation to the true top quark mass. }
   \label{fig:corr_mtop}  
\end{figure}

To reconstruct the top quark mass from the FCNC vertex, three algorithms were developed, as the results  illustrated in 
Figure \ref{fig:corr_mtop}. In the first method (green histogram), the top quark mass is reconstructed using the two opposite-sign (OS) 
electrons with the smallest separation, along with the leading non-$b$-tagged jet. In the second method (red histogram), 
all available electrons and the leading non-$b$-tagged jet are used to minimize the difference $|m_{q\ell\ell} - m_{\rm top}|$. 
The third and optimal algorithm (blue histogram) utilizes all electrons and all non-$b$-tagged jets to minimize $|m_{q\ell\ell} - m_{\rm top}|$, 
producing the most accurate reconstruction of the top quark mass.
All methods yield maximum values around the expected top quark mass, with the blue algorithm demonstrating superior performance in achieving this objective.

\begin{figure}[h]
  \begin{subfigure}{.5\textwidth}
  \centering
    \includegraphics[width=.9\linewidth]{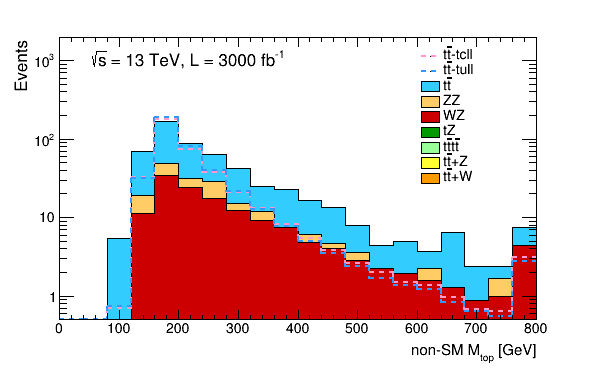}
  \end{subfigure}%
  \begin{subfigure}{.5\textwidth}
  \centering
    \includegraphics[width=.9\linewidth]{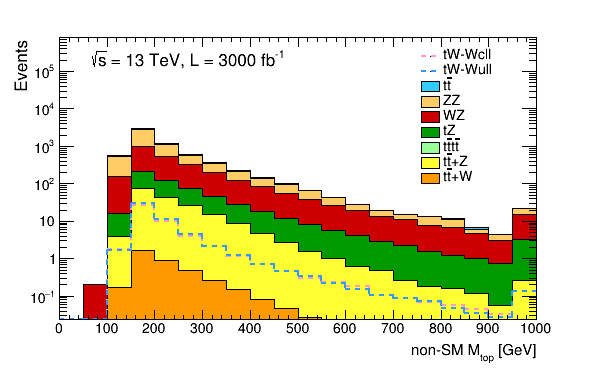}
  \end{subfigure}
  \caption{Distribution of top quark mass from non-standard vertex for $t\bar{t}$ (left) and $tW$ (right) signal scenarios with $S_{RR} = V_{RR} = T_{RR} = 1$ and $\Lambda=1$ TeV, overlaid with the SM backgrounds. The last bin (overflow bin) collects all values exceeding the upper range of the histogram.}
   \label{fig:mtop}  
\end{figure}

Figure \ref{fig:mtop} shows the distribution of the reconstructed top quark mass from the FCNC vertex for both $t\bar{t}$ (left) and $tW$ (right) signal processes. 
The distributions include the main background contributions and reflect the application of the selection criteria described in the previous sections.

\begin{figure}[ht]
  \centering
    \includegraphics[width=0.8\linewidth]{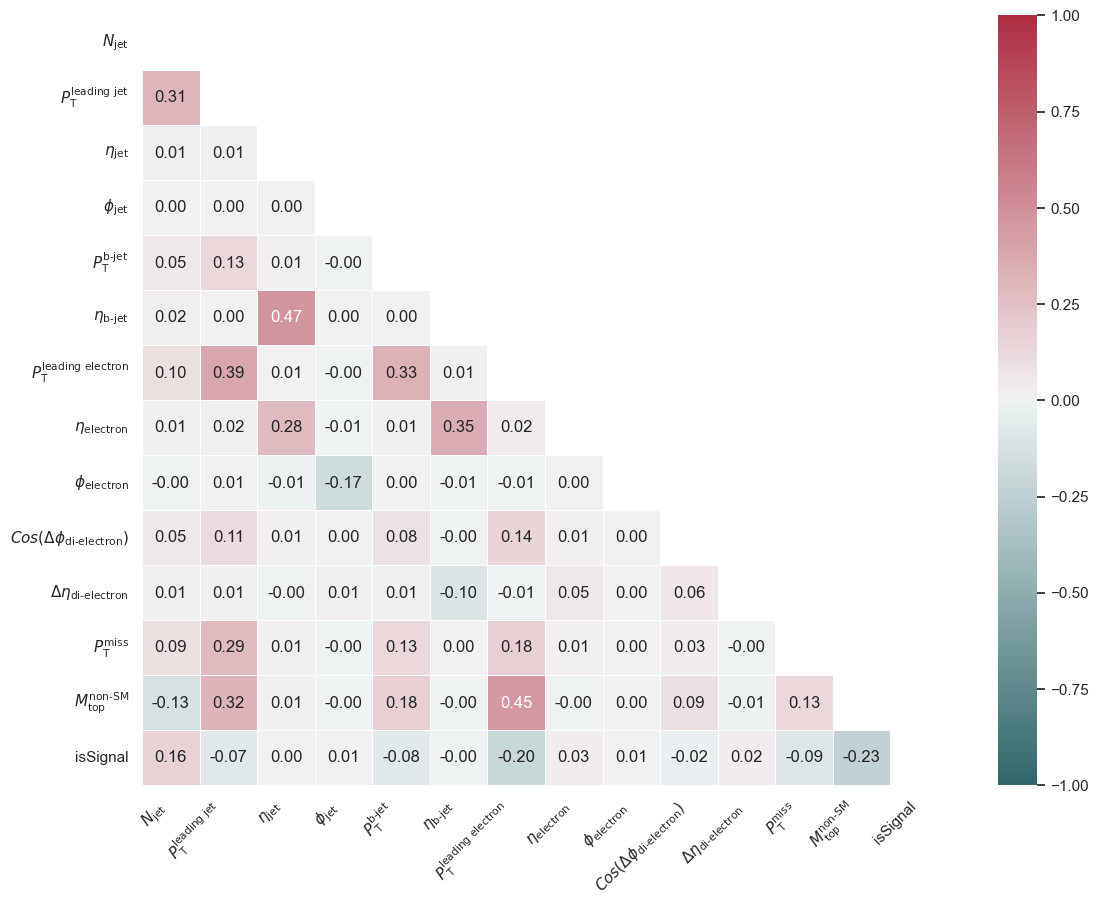}
   \caption{ Heatmap of input feature correlations for the $t \to ue^{-}e^{+}$ decay channel. 
   Each cell represents the correlation coefficient between two variables, highlighting relationships between features. 
   The last row, labeled 'isSignal,' indicates the correlation of each feature with the classification of events as either signal or background. }
   \label{fig:heatmap}  
\end{figure}

In order to discriminate the signal events from the SM backgrounds, several variables were defined as the 
input for the machine learning models listed below: 
\begin{itemize}
\item{$p_{\rm T}, \eta$, and $\phi$ of the leading light flavor jet.}
\item{$p_{\rm T}$ and $\eta$ of the b-tagged jet.}
\item{$p_{\rm T}, \eta$, and $\phi$ of the leading electron.}
\item{cosine of the angle between the first and second opposite sign electrons ($\cos(\Delta\phi(e_1,e_2))$).}
\item{difference between the first and second opposite sign electrons in $\eta$ ($\Delta\eta(e_{1},e_{2})$).}
\item{the magnitude of the missing transverse momentum ($p_{\rm T}^{ \rm miss}$).}
\item{the invariant mass of non-SM top quark, i.e. $m_{eeu}$, indicated as $M_{\rm top}^{\rm non-SM}$.}
\end{itemize}
Figure \ref{fig:heatmap} illustrates the roles of input features heatmaps for $t \to ue^{-}e^{+}$ in this analysis. 
The final row ("isSignal") shows whether each event is classified as signal or background. 
This label is determined during the event generation process, where signal and background samples are simulated separately. 
The most important feature, with the highest correlation coefficients relative to the "isSignal" row, is the non-standard model top quark mass, 
with a coefficient of -0.23. The negative correlation coefficient for the non-standard model top quark mass indicates that as the value 
of the non-standard model top quark mass increases, the probability of the data being background becomes higher.

At the end of this section, we discuss the validity of the EFT framework, 
which requires that contributions from new physics remain within the bounds imposed by tree-level unitarity. 
In particular, new physics induced cross-sections must not exceed these limits to ensure the consistency of the EFT expansion. 
The four-fermion operators considered in this study can arise from different ultraviolet (UV) completions, 
such as a heavy neutral boson mediating interactions between quarks and leptons or a scalar or 
vector particle coupling to quark-lepton pairs. In the case of an intermediate heavy boson, 
the EFT description remains valid as long as the cutoff scale $\Lambda_{e}$ satisfies  
$m_{e^{+}e^{-}}^{\text{max}} ~<  ~\Lambda_{e}$, where $m_{e^{+}e^{-}}^{\text{max}}$ represents the maximal invariant mass of the di-electron (involved in the $tu(c)ee$ vertex) 
in the process. 
This condition is applied to ensure that the characteristic energy of the interaction remains below the new physics scale, 
thereby preserving the validity of the EFT approximation and maintaining the reliability of the effective expansion.

\section{Machine Learning Models}
\label{sec:machinelearning}

The traditional cut-based selection method is a relatively crude approach, 
potentially resulting in the loss of significant signal due to the absolute thresholds imposed. 
Machine learning techniques are keys in the search for flavor-changing processes involving the top quark. 
Methods like de­ep learning and boosted decision trees effectively distinguish between signal events and 
background processes \cite{susystrong} and \cite{susyweak}, improving the sensitivity of experimental searches. 
For the implementation of Neural Netwrok (NN) we used Keras \cite{KERAS}, an open-source library 
acts as an interface for the TensorFlow package. The Scikit-learn \cite{SCIKIT} library was employed for 
various other classification algorithms, including boosted decision trees, random forests, support vector machine, and more.

\subsection{Classification Neural Network}
A neural network consists of interconnected nodes arranged in layers. Each node processes information and transfers it to the next layer.
The links among nodes are determined by a weight, the values of which are adjusted throughout the training process. 
The neural network modifies its weighted connections based on a Loss Function (such as mean squared error, 
which measures the difference between actual and desired outputs) in an effort to minimize the loss value. 
Each iteration where the algorithm processes the entire dataset is termed an epoch, while the extent to which 
weights are adjusted during each epoch is denoted as the learning rate. If the learning rate is excessively high, 
the model may rapidly converge to a suboptimal solution, whereas an overly low learning rate can impede progress, leading to stagnation. 
We employed a set of significant kinematic properties, such as non-SM top quark mass, jet's and electron's transverse momentum, 
missing transverse momentum, etc., as input features. The neural network model was trained using labeled data, 
where Signal instances were assigned a score of 1 and Background instances a score of 0. 
The dataset was partitioned into training and testing sets, with an 80$\%$-20$\%$ split, where the 
test set was utilized to evaluate performance.

Neural network hyperparameter optimization is a critical step in designing efficient and effective deep learning models. 
These hyperparameters, such as learning rate, batch size, number of layers, neurons per layer, optimizes, and activation 
functions influence the performance of the network. Techniques for hyperparameter optimization aim to find the optimal 
combination of hyperparameters that minimize the model's loss function and improve its predictive accuracy. 
In this search we used random search method to explore the hyperparameter space and identify the best configuration. 
To mitigate overfitting in the neural network, two strategies are employed. First, early stopping which is an effective approach 
where the training process is halted when the model's performance on a validation set begins to degrade, preventing it from further 
memorizing noise in the training data. Additionally, dropout can be implemented during training to randomly deactivate a fraction of neurons, 
encouraging the network to learn more robust features and reducing reliance on specific neurons. 

\begin{figure}[h]
	\centering
	\includegraphics[width=.6\linewidth]{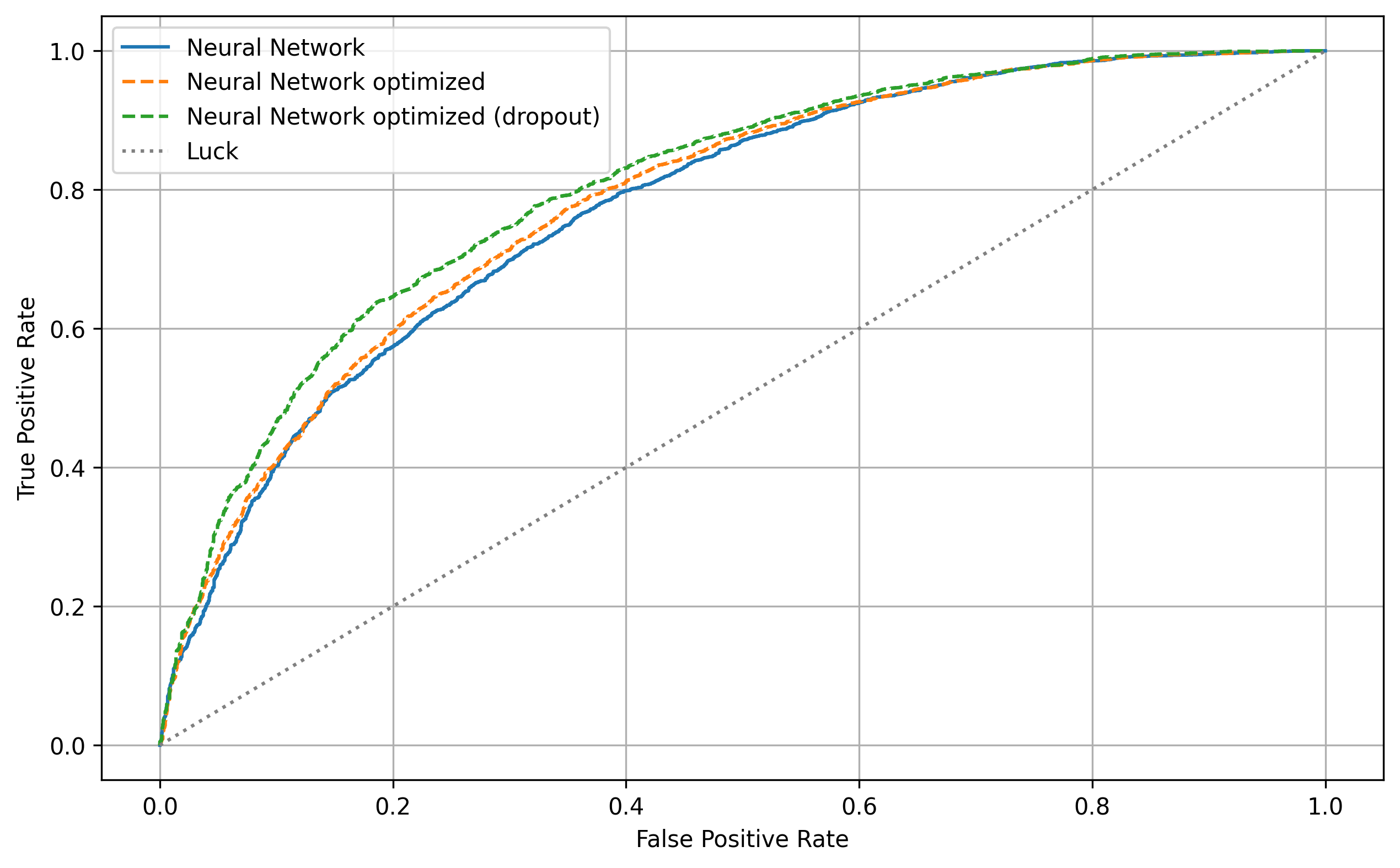}
	\caption{ROC curve showing the performance of the three neural network classification models, with the x-axis representing the False Positive Rate (FPR) and the y-axis representing the True Positive Rate (TPR).}
	\label{fig:roc}
\end{figure}

The effectiveness of a binary neural network, consisting of only two outputs, can be evaluated using the Receiver Operating Characteristic (ROC) curve. 
This curve illustrates the relationship between the true positive rate (signals which are labeled as signals) and the false positive rate 
(backgrounds which are labeled as signals) across different threshold settings, as depicted in Figure \ref{fig:roc}. 
A greater area under the ROC curve generally indicates higher performance of the neural network. 
The neural network equipped with fine-tuned hyperparameters and dropout layers (shown as green in Figure \ref{fig:roc}) achieves the highest performance in detecting signal events.

\begin{figure}[ht]
	\centering
	\includegraphics[width=.6\linewidth]{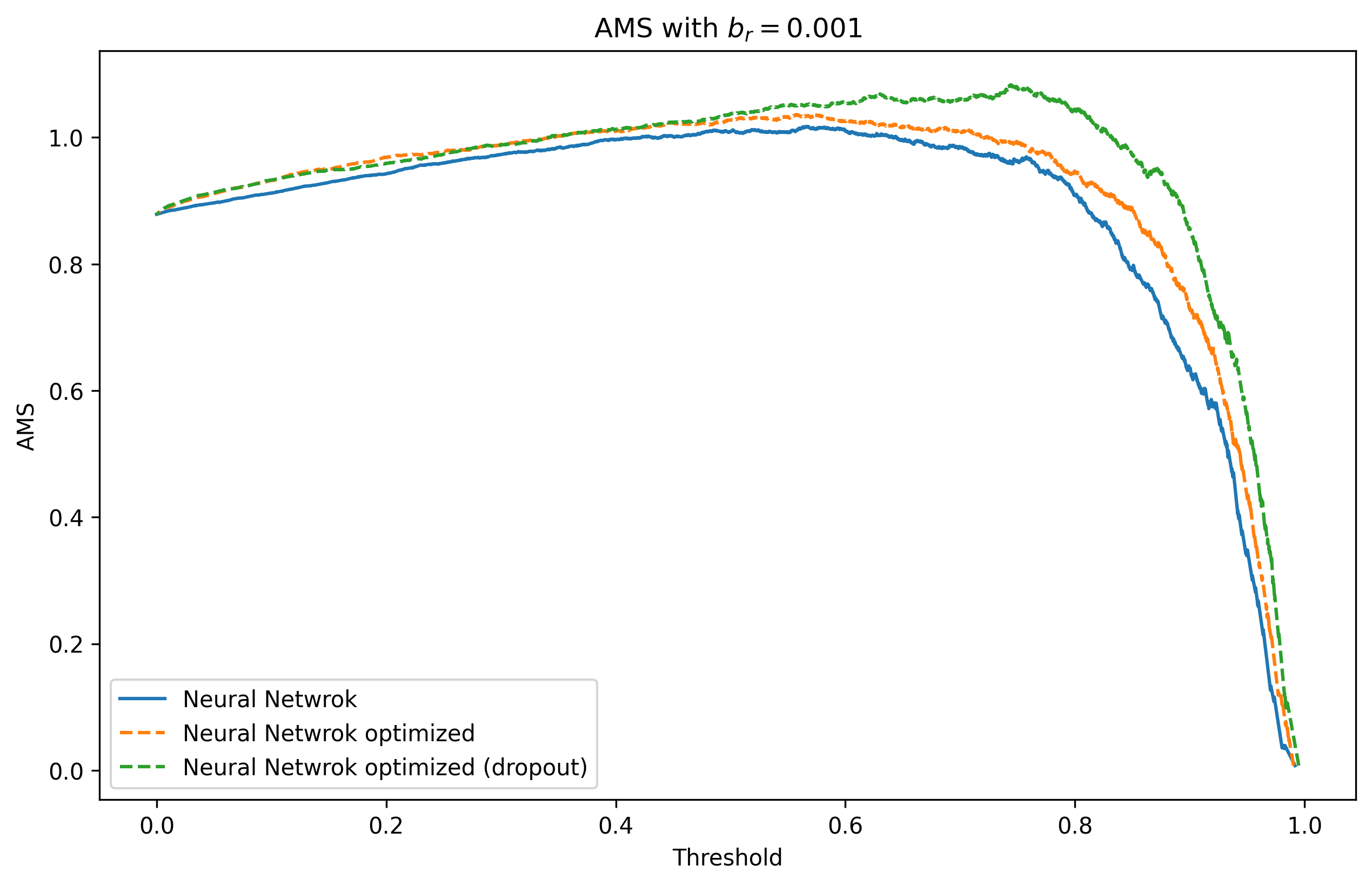}
	\caption{AMS curve for a simple neural network (blue), optimized neural network (orange), and optimized neural network with dropout layers (green) for the $t\bar{t}$ signal. The threshold with the highest AMS value is the best option for defining the signal region.}
	\label{fig:ams}
\end{figure}

To define signal regions, we used approximate median significance (AMS):
\begin{equation}
	\mathrm{AMS}=\sqrt{2\left(\left(TPR+FPR+b_r\right) \ln \left(1+\frac{T P R}{F P R+b_r}\right)-T P R\right)}
\end{equation}

where $TPR$ and $FPR$ are the true and false positive rates and $b_r$ is some number chosen to reduce the variance of the AMS such that the selection region is not too small. The significance of an observed signal is then evaluated based on the value of the AMS. we have plotted AMS for several values of $b_r$ to see how it affects selection for the threshold value and found that $b_r = 0.001$ maximizes the AMS plots. A higher AMS value indicates a greater deviation of the observed signal from the expected background, suggesting a more significant observation, one should then select the value of the threshold that maximizes the AMS. The impact of the threshold on the AMS values for all neural network models for $t\bar{t}$ signal is shown in Figure \ref{fig:ams}. As observed, the optimized neural network with dropout layers exhibits a higher AMS value, particularly around the $70\%-80\%$ threshold.

\subsection{Other classification models}
\begin{figure}[h]
	\centering
	\includegraphics[width=.75\linewidth]{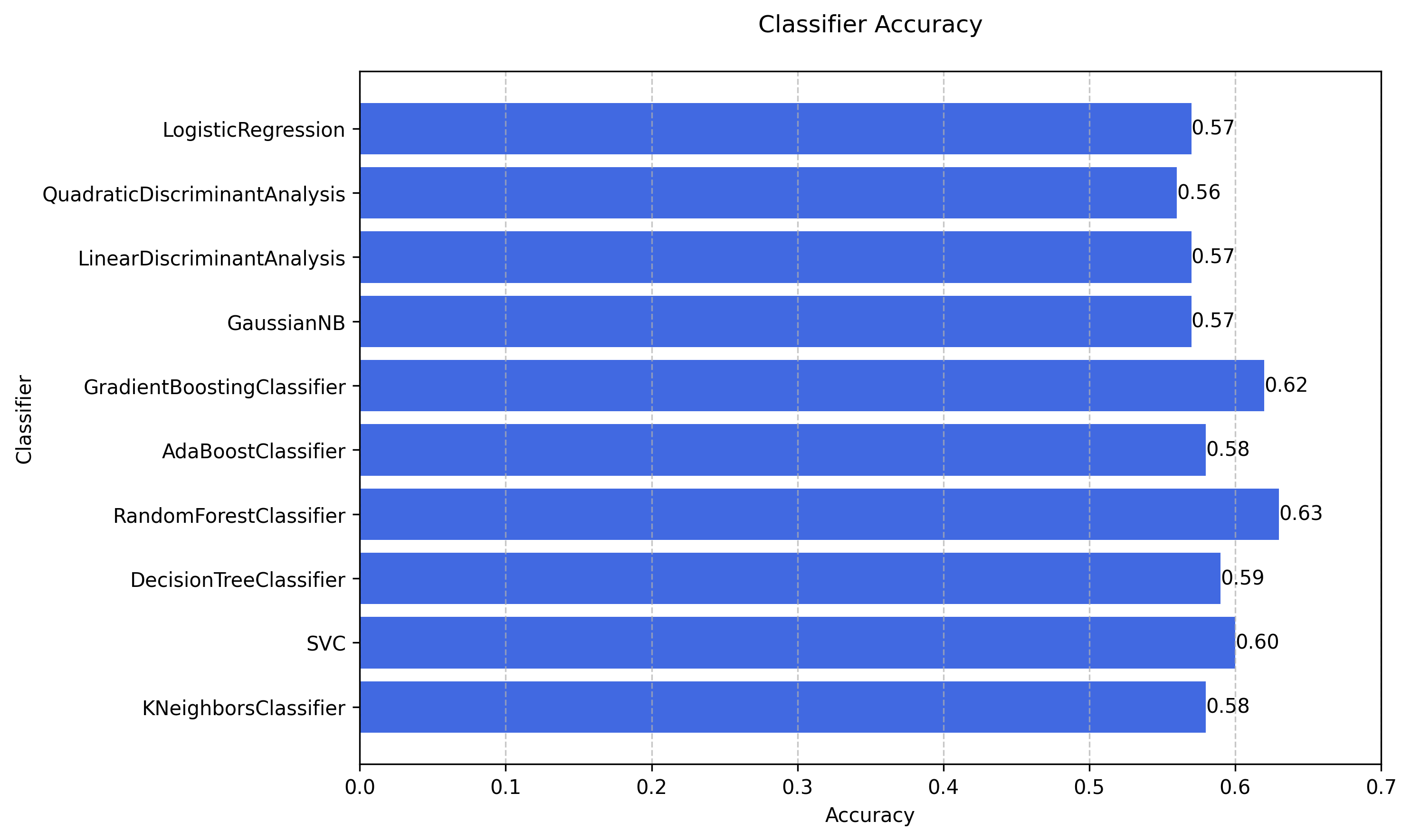}
	\caption{Accuracy for classified models such as random forest, support vector machine, and gradient boosting, etc. The hyperparameters for random forest and decision tree classifiers are optimized to achieve the maximum accuracy.}
	\label{fig:otherml}
\end{figure}

In addition to the neural network model, various other classification machine learning algorithms have been explored in this study. Among these are random forest \cite{RF}, decision tree \cite{BDT}, support vector machine (SVM) \cite{SVM}, k-nearest neighbors (KNN) \cite{KNN}, and simple logistic regression. Each of these models offers unique advantages and characteristics that make them suitable for different types of classification tasks. Random forest and decision tree models, for example, are known for their ability to handle non-linear relationships and interactions between features, making them effective in capturing complex patterns in the data. Support vector machines excel at separating classes with a clear margin of separation, while k-nearest neighbors relies on similarity metrics to classify data points. By exploring a range of classification algorithms, we aim to identify the most suitable model for accurately detecting and characterizing top quark flavor changing events in simulated data. Figure \ref{fig:otherml} illustrates the accuracy of various classification machine learning models across a spectrum of known classifiers. Compared to the $62\%$ and $63\%$ accuracy achieved by the gradient boosting and random forest models, the optimized neural network with dropout layers achieves a higher accuracy of $76\%$.

\section{Results}
\label{res}

In this sensitivity analysis, our objective is to evaluate the sensitivity of the experimental results to the scale of new physics, 
represented by the parameter $\Lambda$. We begin by calculating the p-value for each hypothesis, 
both in the presence and absence of signal events, using the BinomialExpP function within the RooFit 
framework \cite{ROOFIT}. This approach allows us to quantify the probability of observing the expected 
event distribution under each hypothesis, based on simulated data.
 To further interpret the results, we perform a CLs test, a statistical method widely adopted 
in particle physics analyses, to establish the $95\%$ Confidence Level (CL) exclusion thresholds for 
the parameter $\Lambda$. These exclusion thresholds define the range of $\Lambda$ values excluded 
at the $95\%$ CL, providing constraints on the scale of new physics under investigation.

In Tables \ref{tttab} and \ref{twtab}, we present the expected number of signal and total background events 
per $3000$ fb$^{-1}$ of integrated luminosity for two distinct signals, $t\bar{t}$ and $tW$, 
where the top quark decays to a light-flavor jet ($u$ or $c$) and an electron-positron pair. 
Results are listed for the optimized neural network model with dropout layers (the most accurate model), 
along with two threshold settings that vary depending on the signal scenario.
As observed, increasing the cut on the neural network (NN) weights leads to a 
higher signal-to-background ratio ($\mathcal{N}_{S}$), enhancing the separation between signal and background. 
This improvement allows for greater sensitivity in distinguishing potential signal events from background noise, 
contributing to more robust constraints on the parameter $\Lambda$.

\begin{table}[h!]
\centering
\caption{\small Number of signal ($\mathcal{N}_{S}$), background ($\mathcal{N}_{B}$) events, 
and signal-to-background ratio per 3000 fb$^{-1}$ of integrated luminosity, for the process $pp \to t\bar{t} \to u e^{-}e^{+} \bar{t}$ (top) and $pp \to t\bar{t} \to c e^{-}e^{+} \bar{t}$ (bottom).}\label{tttab}
\begin{tabular}{c|c|c|c|c|c|c}
\toprule
& \multicolumn{3}{c|}{NN weights $\geq 75\%$} & \multicolumn{3}{c}{NN weights $\geq 80\%$} \\
\cmidrule(lr){2-4} \cmidrule(lr){5-7}
Process & $\mathcal{N}_{S}$ & $\mathcal{N}_{B}$ & $\mathcal{N}_{S}/\mathcal{N}_{B}$ & $\mathcal{N}_{S}$ & $\mathcal{N}_{B}$ & $\mathcal{N}_{S}/\mathcal{N}_{B}$ \\
\midrule
\multirow{2}{*}{$pp \to t\bar{t} \to u e^{-}e^{+} \bar{t}$} & 538.1 & 345.6 & 1.6 & 421.0 & 245.1 & 2.2 \\
 & & & & & & \\
\midrule
\multirow{2}{*}{$pp \to t\bar{t} \to c e^{-}e^{+} \bar{t}$} & 570.8 & 240.7 & 2.4 & 492.8 & 173.9 & 3.3 \\
 & & & & & & \\
\bottomrule
\end{tabular}
\end{table}

\begin{table}[h!]
\centering
\caption{\small Number of signal ($\mathcal{N}_{S}$), background ($\mathcal{N}_{B}$) events, 
and signal-to-background ratio per 3000 fb$^{-1}$ of integrated luminosity, for the process $pp \to tW \to u e^{-}e^{+} W$ (top) and $pp \to tW \to c e^{-}e^{+} W$ (bottom).}\label{twtab}
\begin{tabular}{c|c|c|c|c|c|c}
\toprule
& \multicolumn{3}{c|}{NN weights $\geq 65\%$} & \multicolumn{3}{c}{NN weights $\geq 70\%$} \\
\cmidrule(lr){2-4} \cmidrule(lr){5-7}
Process & $\mathcal{N}_{S}$ & $\mathcal{N}_{B}$ & $\mathcal{N}_{S}/\mathcal{N}_{B}$ & $\mathcal{N}_{S}$ & $\mathcal{N}_{B}$ & $\mathcal{N}_{S}/\mathcal{N}_{B}$ \\
\midrule
\multirow{2}{*}{$pp \to tW \to u e^{-}e^{+} W$} & 47.3 & 4087.4 & 0.013 & 38.7 & 2955.1 & 0.019 \\
 & & & & & & \\
\midrule
\multirow{2}{*}{$pp \to tW \to c e^{-}e^{+} W$} & 38.1 & 3353.7 & 0.014 & 30.1 & 2261.9 & 0.020 \\
 & & & & & & \\
\bottomrule
\end{tabular}
\end{table}

\begin{figure}[h]
  \begin{subfigure}{.5\textwidth}
  \centering
    \includegraphics[width=.9\linewidth]{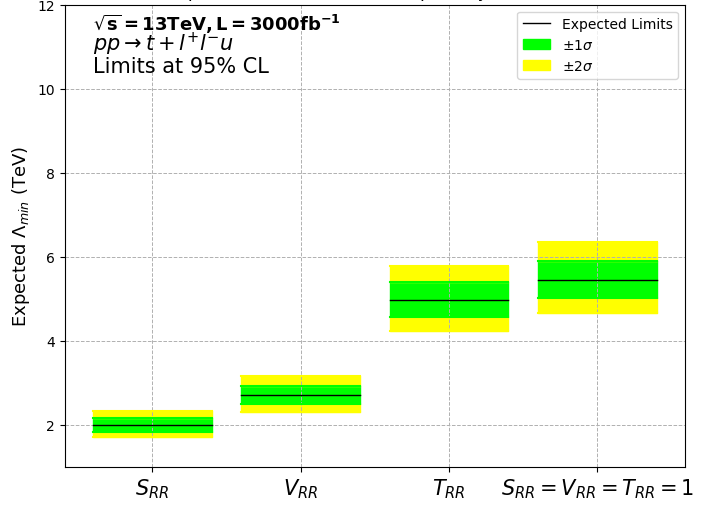}
  \end{subfigure}%
  \begin{subfigure}{.5\textwidth}
  \centering
    \includegraphics[width=.9\linewidth]{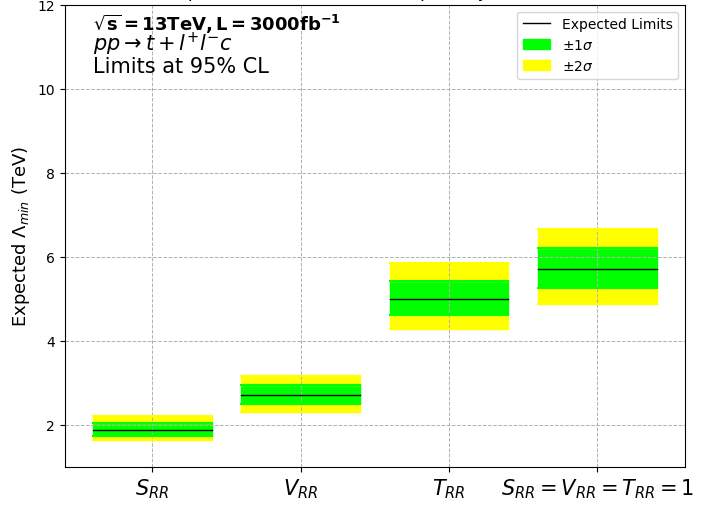}
  \end{subfigure}
  \caption{Expected 95\% CL lower limit on $\Lambda_{e}$, of the $tuee$ (left) and $tcee$ (right) for 4 signal scenarios: $S_{RR} = 1$, $V_{RR} = 1$, $T_{RR} = 1$, and $S_{RR} = V_{RR} = T_{RR} = 1$, and total integrated luminosity of 3000 fb$^{-1}$.}
   \label{fig:cls}  
\end{figure}

The expected $95\%$ CL exclusion limits on the new physics scale $\Lambda$ for the $t\bar{t}$ process are shown in 
Figure \ref{fig:cls}, for three coupling scenarios: scalar-like ($S_{RR} = 1$), vector-like ($V_{RR} = 1$), tensor-like ($T_{RR} = 1$), 
and a mixed scenario ($S_{RR} = V_{RR} = T_{RR} = 1$). 
These results are obtained using the optimized neural network model with dropout layers, and are presented 
alongside $\pm1\sigma$ and $\pm2\sigma$  uncertainty bands. 
Among these scenarios, the scalar-like configuration demonstrates a lower exclusion limit,  due to its reduced production cross-section. 
The sensitivity of the $t\bar{t}$ channel to tensor (vector) coupling extends to a lower limit of approximately $\Lambda_{e} > 5 (2.8)$ TeV for both $tuee$
and $tcee$ at an integrated luminosity of $3000$ fb$^{-1}$.
The sensitivity achieved in the analysis of single top plus dilepton to tensor (vector) was found to be $\Lambda_{\mu} > 7.1 (4.7)$ TeV  and $\Lambda_{\mu} >  2.4 (1.5)$ TeV 
 for the corresponding $tu\mu\mu$ and $tc\mu\mu$ using an integrated luminosity of $3000$ fb$^{-1}$ \cite{p4}.

\begin{figure}[h]
  \begin{subfigure}{.5\textwidth}
  \centering
    \includegraphics[width=.9\linewidth]{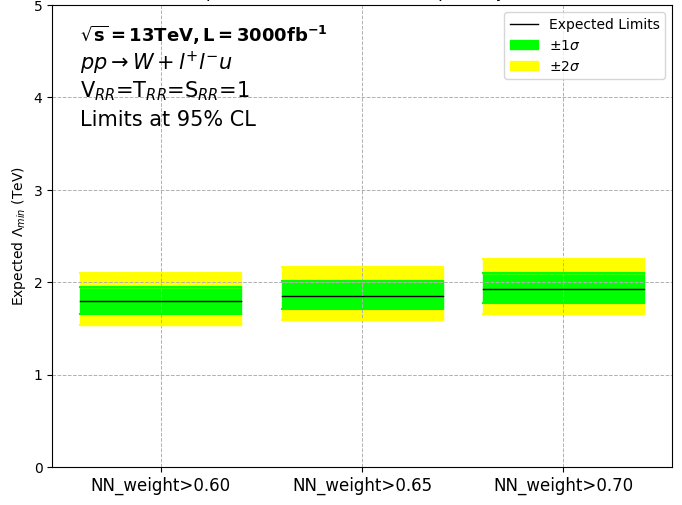}
  \end{subfigure}%
  \begin{subfigure}{.5\textwidth}
  \centering
    \includegraphics[width=.9\linewidth]{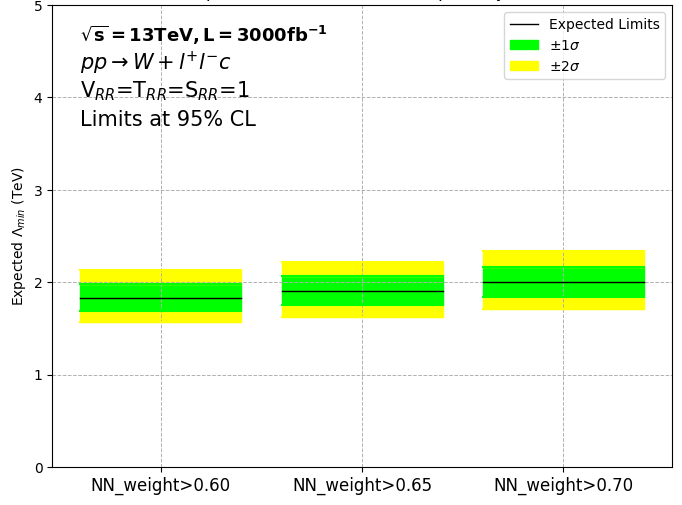}
  \end{subfigure}
  \caption{Expected 95\% CL lower limit on $\Lambda_{e}$, of the $tuee$ (left) and $tcee$ (right) of the $tW$ process for three neural network thresholds and total integrated luminosity of 3000 fb$^{-1}$.}  
   \label{fig:twcls}  
\end{figure}

Figure \ref{fig:twcls} shows the corresponding $95\%$ CL exclusion limits on $\Lambda_{e}$ for 
the $tW$ process across three distinct neural network threshold settings, all utilizing the
 optimized network with dropout layers. The results indicate consistent performance across thresholds, 
 with lower exclusion limits reaching up to 2 TeV. 
 These findings underscore the robustness of the analysis across both processes, 
 demonstrating competitive sensitivity in probing the scale of new physics in FCNC interactions.

\section{Interpretation of Results in the Context of a \(Z'\) Model}
\label{sec:int}

In this section, as an example we interpret the results of our EFT analysis in the context 
of a specific \(Z'\) model. The \(Z'\) boson is a new gauge boson associated with an additional 
\(U(1)'\) gauge symmetry beyond the SM. 
The Lagrangian describing the interactions of the \(Z'\) boson with the SM fermions is given by:
\begin{eqnarray} \label{zplag}
\mathcal{L} \supset g_{tu} Z'_\mu \bar{t} \gamma^\mu P_L u + g_\ell Z'_\mu \bar{\ell} \gamma^\mu P_L \ell + \text{h.c.},
\end{eqnarray}
where \(Z'_\mu\) is the \(Z'\) gauge field, \(g_{tu}\) and \(g_\ell\) are the couplings of the \(Z'\) boson 
to the top quark (\(t\)), up/charm quark (\(u\)), and leptons (\(\ell\)), respectively.
 This model introduces FCNCs at tree level and the \(Z'\) boson 
can mediate processes such as \(t \to u Z'\) and \(Z' \to \ell^+ \ell^-\). By matching the EFT to this \(Z'\) model, 
we can translate the bounds on the EFT Wilson coefficients into constraints on the \(Z'\) couplings and mass. 
This interpretation allows us to explore the implications of our EFT results for new physics scenarios involving a \(Z'\) boson with flavor-violating interactions.
In particular, the matching condition is given by:  $\frac{V}{\Lambda^2} \propto  g_{tu} g_\ell/M_{Z'}^2$.

\begin{figure}[ht]
	\centering
	\includegraphics[width=0.7\linewidth]{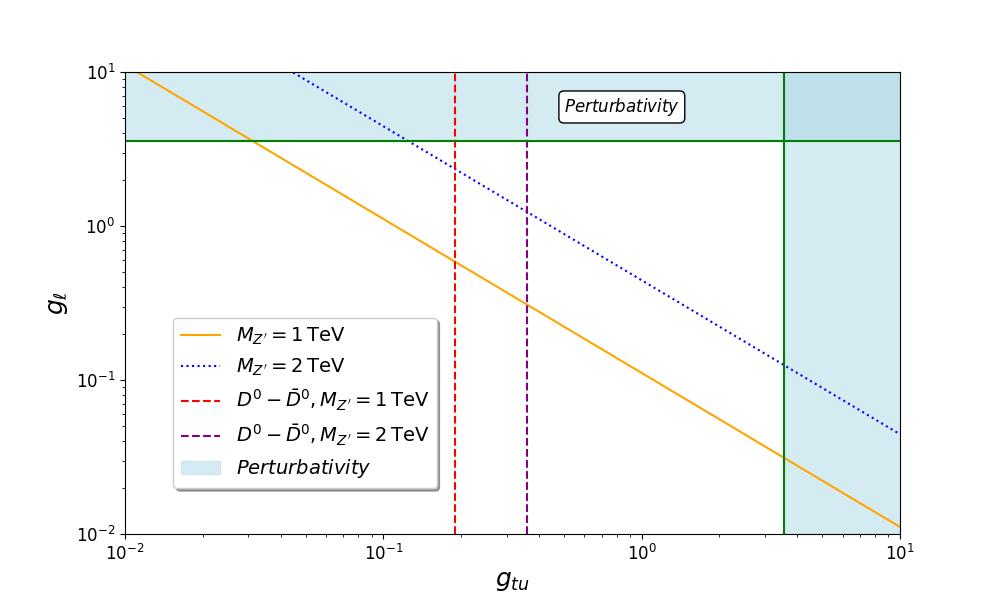}
	\caption{Parameter space constraints derived from the interpretation of rare top decays at the HL-LHC for \(Z'\) boson masses of \(m_{Z'} = 1\) and \(2\) TeV. The light blue shaded region represents the exclusion derived from perturbativity conditions. The dashed lines are the limits obtained from $D^{0}-\bar{D}^{0}$ mixing assuming $g_{tc} = 0.01$ for  \(m_{Z'} = 1\) and \(2\) TeV. }
	\label{fig:zp}
\end{figure}

This matching condition arises from integrating out the heavy \(Z'\) boson and expressing 
its effects in terms of higher-dimensional operators in the EFT. The Wilson coefficient \(V\) 
encapsulates the effects of the \(Z'\) boson at energies below its mass scale, allowing us to 
connect the EFT framework to the specific UV completion provided by the \(Z'\) model.
The matching condition plays a crucial role in translating the bounds on the Wilson coefficient, 
obtained from experimental analysis and EFT analyses, into constraints on the fundamental parameters 
of the \(Z'\) model: the couplings \(g_{tu}\) and \(g_\ell\), and the \(Z'\) mass \(m_{Z'}\). 
Figure \ref{fig:zp} shows the parameter space region that can be bounded by the interpretation of the the present analysis of 
rare top decays in the HL-LHC with $m_{Z'} = 1, 2$ TeV.

In addition to experimental constraints, the \(Z'\) model must satisfy theoretical requirements, 
such as perturbativity, to ensure its consistency as a quantum field theory.
The decay $t \to u+ e^{+} + e^{-}$ is mediated by the $Z'$ boson, and its amplitude depends on 
the product of the couplings $g_{tu}$ and $g_\ell$. To ensure perturbativity, the following conditions must hold:
$|g_{tu}| \leq \sqrt{4\pi} \quad \text{and} \quad |g_\ell| \leq \sqrt{4\pi}$.
These conditions ensure that the perturbative expansion of the decay amplitude remains valid, 
preventing higher-order terms from dominating the leading-order results and preserving the consistency of the theory. 
The constraints derived from these conditions are illustrated in Figure \ref{fig:zp}, where the excluded region is highlighted in light blue shading.

We also present limits on the flavor-changing part of the \( Z' \) couplings using \( D^0-\bar{D^0} \) 
mixing, considering its effects on the up-type quarks and constraining the couplings and mass of \( Z' \).
Within the SM, \( D^0-\bar{D^0} \) mixing is a manifestation of FCNCs that occurs because the flavor eigenstates 
differ from the physical mass eigenstates of the \( D^0-\bar{D^0} \) system. 
Both short-range quark-level transitions and long-range processes contribute to \( D^0-\bar{D^0} \) oscillations. 
The short-range contributions proceed through loops where virtual particles are mediated \cite{Bianco:2003vb, Burdman:2003rs, Artuso:2008vf}. 
This makes the study of \( D^0-\bar{D^0} \) mixing particularly interesting, 
as new physics models with additional FCNC degrees of freedom can be examined through it.
The parameters of the $Z'$ FCNC interactions between two up-type quarks can be significantly 
constrained using the measurement of \( D^0-\bar{D^0} \) mixing, which is affected at both the tree level and loop level. 
Assuming the mediator has up-charm (\( g_{uc} \)), up-top (\( g_{tu} \)), 
and charm-top (\( g_{tc} \)) couplings, it can have a significant contribution to neutral \( D \) meson mixing. 
If the mediator is \( Z' \), its contribution to the mass difference between the two mass eigenstates is given by \cite{Arhrib:2006sg}:
\begin{align}
\Delta M_D &= \frac{f_D^2 M_D^2 B_D}{48 m_{Z'}^2} \left[ g_{uc}^{2} + g_{tu}^{2} g_{tc}^{2} \frac{x}{8 \pi^2} \left(32 f_{_{Z'}}(x) - 5 g_{_{Z'}}(x)\right) \right], \\
\text{where } & f_{_{Z'}}(x) = \frac{1}{2} \frac{1}{(1-x)^3} \left[1 - x^2 + 2x \log x\right], \nonumber \\
& g_{_{Z'}}(x) = \frac{2}{(1-x)^3} \left[2(1 - x) + (1 + x)\log x\right]. \nonumber
\end{align}
The first term in the square bracket represents the tree-level contribution, 
while the second term is due to the box contribution. 
Here, \( x \equiv m_{Z'}^2/m_t^2 \), \( M_D \sim 1.9\ \text{GeV} \) is the mass of the \( D \) meson, 
\( f_D \sim 0.223\ \text{GeV} \) is its decay constant, \( B_D \sim 1 \) is the bag model parameter. 
In this work, for simplicity we set $g_{uc} = 0$ and only concentrate on the FCNCs of top and up/charm quarks via \( Z' \).
The current bound on \( D \) meson mixing is $\Delta M_D \lesssim  10^{-15}$~\cite{dm1,dm2}, 
which allows us to derive constraints on $g_{tu}$, as presented in Fig.~\ref{fig:zp} for 
$g_{tc} = 0.01$ and $m_{Z'} = 1,2$ TeV. The bounds from $D$ meson mixing on $g_{tu}$ and $g_{tc}$ are similar.

\section{Summary}
\label{summ}

In this study, we have investigated the effects of FCNC couplings in the top-quark sector through the processes 
$pp \to t\bar{t}$ and $pp \to tW$, effectively probed using a three-electron selection and a $b$-jet tagging strategy. 
SM backgrounds were carefully analyzed, demonstrating that an optimal selection criterion based on the 
reconstructed top quark mass from the FCNC vertex provides excellent separation between 
NP signals and SM backgrounds. Several machine learning models were trained and optimized to enhance 
signal-background discrimination, with the neural network incorporating dropout 
layers yielding the highest performance in classification metrics. Furthermore, 
we calculated exclusion limits on the scale of the $tuee$ and $tcee$ FCNC interactions, with expected 
$95\%$ CL bounds of $\Lambda_{e} \leq 5.5$ TeV for a $tuee$ interaction and $\Lambda_{e} \leq 5.7$ TeV 
for a $tcee$ interaction, assuming $S_{RR} = V_{RR} = T_{RR} = 1$.
For the only tensor interaction, the expected $95\%$ CL bounds of $\Lambda_{e} \lesssim 5$ TeV 
for both $tuee$ and $tcee$ interactions have obtained.

These results can be compared to the recent study \cite{p4} that explores
FCNC interactions in $tu\mu^{+}\mu^{-}$ and $tc\mu^{+}\mu^{-}$ processes, based on 3000 fb$^{-1}$ of HL-LHC data, which
reports $95\%$ CL bounds of $\Lambda_{\mu} \leq 7.1$ TeV and $\Lambda_{\mu} \leq 2.4$ TeV for the tensor coupling via
these trilepton signals. 
Despite the lepton flavor difference, our analysis achieves comparable sensitivity, with the expected limit for the $tce^{+}e^{-}$ 
channel reaching a value of around 2.5 TeV higher. 

This study enhances FCNC sensitivity at the LHC by applying machine learning techniques to three-electron final states.
The exclusion limits derived here are competitive with previous analyses, 
highlighting the power of advanced classification models in enhancing new physics searches.
In addition, we have interpreted our EFT results obtained from the analysis in the context of a specific $Z'$ model, 
demonstrating how the derived constraints map onto the parameter space of a heavy neutral gauge boson that mediates flavor-changing interactions.

\section{Acknowledgments}
We are grateful to Gh. Haghighat for his insightful comments on the manuscript. 
We would like to express our sincere gratitude to IPM for 
providing the servers and computational resources necessary for this research.



\end{document}